\pdfoutput=1

\documentclass[11pt]{article}

\usepackage{ACL2023}

\usepackage{times}
\usepackage{latexsym}
\usepackage{booktabs}
\usepackage{multirow}
\usepackage{graphicx}
\usepackage{bbm}
\usepackage[acronym]{glossaries}
\usepackage{acronym}
\usepackage{enumitem}
\usepackage{subcaption}
\usepackage{titlesec}
\usepackage{tikz}
\usepackage{tcolorbox}
\usepackage{lipsum,multicol}
\setlist[itemize]{leftmargin=*}

\newtcolorbox[auto counter,number within=section]{pabox}[2][]{%
colback=red!5!white,colframe=red!75!black,fonttitle=\bfseries,
title=Prompt.~\thetcbcounter: #2,#1}

\titleformat{\subsubsection}[runin]{\bfseries}{}{0em}{}[]

\acrodef{LSR}{Learned Sparse Retrieval}
\acrodef{DR}{Dense Retrieval}
\acrodef{BOWP}{bag of word pieces}
\acrodef{MLM}{Masked Language Model}
\acrodef{LLMs}{Large Language Models}
\usepackage[T1]{fontenc}

\usepackage[utf8]{inputenc}

\usepackage{microtype}

\usepackage{inconsolata}

\title{DyVo: Dynamic Vocabularies for Learned Sparse Retrieval with Entities}
\author{
\textbf{Thong Nguyen\textsuperscript{1}}, 
\textbf{Shubham Chatterjee\textsuperscript{2}},
\textbf{Sean MacAvaney\textsuperscript{3}} \\
\textbf{Iain Mackie\textsuperscript{3}},
\textbf{Jeff Dalton\textsuperscript{2}},
\textbf{Andrew Yates\textsuperscript{1}}
\\
 \textsuperscript{1}University of Amsterdam,
 \textsuperscript{2}University of Edinburgh,
 \textsuperscript{3}University of Glasgow
\\
 \small{
   \textbf{Correspondence:} \href{mailto:t.nguyen2@uva.nl}{t.nguyen2@uva.nl}
 }
}
\begin{document}
\maketitle
\begin{abstract}
    Learned Sparse Retrieval (LSR) models use vocabularies from pre-trained transformers, which often split entities into nonsensical fragments. Splitting entities can reduce retrieval accuracy and limits the model's ability to incorporate up-to-date world knowledge not included in the training data. In this work, we enhance the LSR vocabulary with Wikipedia concepts and entities, enabling the model to resolve ambiguities more effectively and stay current with evolving knowledge. Central to our approach is a Dynamic Vocabulary (DyVo) head, which leverages existing entity embeddings and an entity retrieval component that identifies entities relevant to a query or document.
    We use the DyVo head to generate entity weights, which are then merged with word piece weights to create joint representations for efficient indexing and retrieval using an inverted index. In experiments across three entity-rich document ranking datasets, the resulting DyVo model substantially outperforms state-of-the-art baselines.\footnote{Code: \url{https://github.com/thongnt99/DyVo}} 
\end{abstract}

\section{Introduction}
\label{sec:Introduction}
Neural IR methods typically operate in two stages. Initially, a set of candidate documents is retrieved using a fast, computationally-efficient first-stage retrieval method that considers sparse or dense vector representations. These candidates are then re-ranked using more computationally-intensive scoring functions, such as those involving cross-encoders~\cite{nogueira2019passage,macavaney2019cedr,nogueira-etal-2020-document,sun-etal-2023-chatgpt}.

\ac{LSR}~\cite{nguyen2023unified, formal2021splade, formal2022distillation} is a prominent neural method for first-stage retrieval. \ac{LSR} encodes queries and documents into sparse, lexically-aligned representations that can be stored in an inverted index for fast retrieval.
%
LSR offers several advantages over Dense Retrieval (DR), another common approach for first-stage retrieval~\cite{lin2020pretrained}. LSR’s lexically grounded representations are more transparent, making it easier for users to understand the model and inspect representations for biases~\cite{abolghasemi2024measuring}.
Furthermore, LSR's compatibility with an inverted index enables efficient and exact retrieval~\cite{DBLP:conf/sigir/DingS11}, while also simplifying the transition from existing lexical search infrastructure supporting methods like BM25. LSR not only performs competitively with DR in terms of performance within the same domain, but it also tends to generalize better across different domains and tasks~\cite{formal2021splade}.


\textbf{\begin{figure}[t]
    \centering
    \includegraphics[trim={2.5cm 10.5cm 2.3cm 10cm},clip, width=0.9\linewidth]{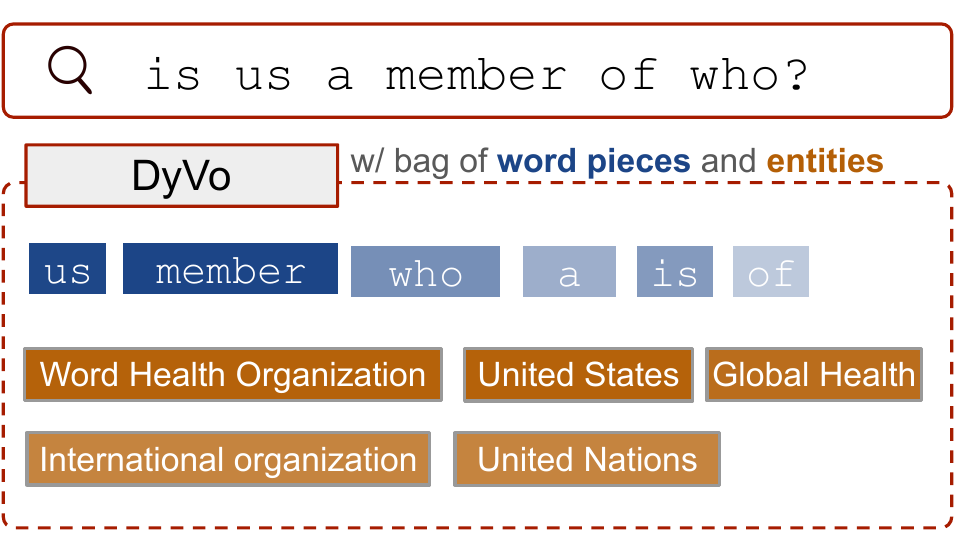}
    \caption{DyVo augments BERT's word piece vocabulary with an entity vocabulary to help disambiguate a query (or document). Word pieces are in blue and entities are in orange. Darker terms have a higher weight in the sparse representation.}
    \label{fig:lsr_example}
    \vspace{-0.4cm}
\end{figure}}

However, LSR models lack explicit representations for entities and concepts in their vocabulary. This can pose challenges due to the tokenization process, where words are segmented into subwords or wordpieces. For instance, a word like ``BioNTech'' might be tokenized into \text{[\textit{bio, \#\#nte, \#\#ch}]}. Such fragmentation can lead to ambiguity, complicating the retrieval process by obscuring the full meaning and context of the original word, which in turn may affect the accuracy and relevance of search results. Additionally, the bag of word pieces representation employed by LSR methods struggles with homonyms, where different meanings or entities, such as ``WHO'' (World Health Organization) and ``US'' (United States), could be conflated when represented merely as word pieces in a query like ``Is the US a member of WHO?''

Hence, while LSR provides a framework for efficient first-stage document retrieval, its current design -- particularly in handling entities and complex vocabulary -- poses significant challenges. We hypothesize that integrating explicit entities into the LSR vocabulary could significantly enhance its performance. 
This integration is especially pertinent as a large proportion of queries pertain to specific entities or are closely related to them ~\cite{kumar2010characterization, guo2009named}. Previous work indicates that hybrid models combining word and entity representations have improved both sparse
 retrieval ~\cite{dalton2014entity, shehata2022early, mackieadaptive2024} and dense retrieval ~\cite{xiong2017word,tran2022dense, chatterjee2024dreq}. 

To address the above limitations, we incorporate entities from Wikipedia into the vocabulary of an LSR model. The English Wikipedia contains entities spanning a diverse range of categories and disciplines, including named entities like people, organizations, and locations, as well as general concepts such as \textit{eudaimonia}, \textit{hot dog}, and \textit{net income}. Integrating these Wikipedia entities into a LSR model significantly enhances its ability to handle complex semantic phrases and entities that are currently fragmented into nonsensical word pieces. By enriching query and document representations with relevant entities, we reduce ambiguity and improve the representational power of LSR. This approach is illustrated in Figure \ref{fig:lsr_example}. Moreover, leveraging Wikipedia -- a rich and continually updated knowledge base -- allows the LSR model to refresh its internal knowledge, aligning it with evolving global information. 

As of April 2024, the English Wikipedia hosts nearly 7 million entities and concepts, which is more than 200 times larger than the word piece vocabulary used in current state-of-the-art LSR methods. To identify relevant entities from among millions of them, we propose adding a Dynamic Vocabulary (DyVo) head with an entity candidate retrieval component.
Specifically, we leverage entity retrieval techniques and Large Language Models (LLMs) to dynamically generate relevant entities. These methods aim to refine the set of highly relevant entities, which are then passed to the \ac{LSR} encoder for scoring. The encoder outputs a small bag of weighted entities, ignoring those that were not retrieved. The entity representation is then concatenated with the word-piece representation, forming a joint representation used for indexing and retrieval processes.

Our contributions are:
\begin{itemize}

\item We propose the DyVo model to address the limitations of the word piece vocabulary commonly employed in \ac{LSR}, which uses a Dynamic Vocabulary (DyVo) head to extend \ac{LSR} to a large vocabulary (e.g., millions of Wikipedia entities and concepts) by leveraging existing entity embeddings and a candidate retrieval component that identifies a small set of entities to score.

\item We introduce a few-shot generative entity retrieval approach capable of generating highly relevant entity candidates, which leads to superior performance when integrated into our DyVo framework.
Furthermore, we find that document retrieval effectiveness using candidates generated by Mixtral or GPT4 is competitive with using entities identified by human annotators.

\item We demonstrate that incorporating entities into \ac{LSR} through a dynamic vocabulary consistently enhances the effectiveness of \ac{LSR} across three entity-rich benchmark datasets (i.e., TREC Robust04, TREC Core 2018, and CODEC). Despite its simplicity, Wikipedia2Vec is a surprisingly effective source of entity embeddings. We achieve further performance gains by utilizing transformer-based dense entity encoders to encode entity descriptions into embeddings.
\end{itemize}

\section{Related Work}
\label{sec:Related Work}

\paragraph{Learned sparse retrieval.} LSR encodes queries and documents into sparse lexical vectors, which are bag of words representations that are indexed and retrieved using an inverted index, akin to traditional lexical retrieval methods like BM25. One of the early works in this area proposed using neural networks to learn sparse representations that are compatible with an inverted index and demonstrated promising performance~\cite{zamani2018neural}. With the advent of the transformer architecture \cite{vaswani2017attention}, subsequent work has successfully utilized pretrained transformers to enhance the effectiveness and efficiency of LSR models \cite{formal2021splade, lassance2022efficiency,formal2022distillation, macavaney2020expansion, zhao2020sparta, zhuang2021tilde}. Among these, SPLADE \cite{formal2021splade, formal2022distillation} stands out as a state-of-the-art LSR method.
While SPLADE uses a word piece vocabulary, prior work has demonstrated that its vocabulary can be replaced by performing additional masked language modeling (MLM) pretraining and then exhaustively scoring all terms in the new vocabulary~\cite{dudek2023learning}.
In this work, we dynamically augment a word piece vocabulary using pre-existing embeddings rather than performing additional pretraining.
SPLADE typically employs a shared MLM encoder for both queries and documents, enabling term expansion and weighting on both sides. However, previous work \cite{nguyen2023unified, macavaney2020expansion} has shown that removing query expansion by replacing the MLM query encoder with an MLP encoder can simplify training and improve efficiency by reducing the number of query terms involved. While most LSR research has focused on ad-hoc paragraph retrieval tasks, recent efforts have explored extending LSR to other settings, such as conversational search \cite{hai2023cosplade}, long documents \cite{nguyen2023adapting}, and text-image search \cite{zhao2023lexlip, chen2023stair, nguyen2024multimodal}.

\paragraph{Entity-oriented search.}  Early work in entity-oriented search primarily utilized entities for query expansion. A significant advancement in this domain was made by \citet{meij2010conceptual}, who introduced a double translation process where a query was first translated into relevant entities, and then the terms associated with these entities were used to expand the query. \citet{dalton2014entity} further developed this concept with Entity Query Feature Expansion, which enhanced document retrieval by enriching the query context with entity features.

The field then recognized the more integral role of entities in search applications, transitioning from merely using entities for query expansion to treating them as a latent layer while maintaining the original document and query representations. Among these methods, Explicit Semantic Analysis \cite{gabrilovich2009wikipedia} used ``concept vectors'' from knowledge repositories like Wikipedia to generate vector-based semantic representations. The Latent Entity Space model \cite{liu2015latent} utilized entities to assess relevance between documents and queries based on their alignments in the entity-informed dimensions. EsdRank \cite{xiong2015esdrank} leveraged semi-structured data such as controlled vocabularies and knowledge bases to connect queries and documents, pioneering a novel approach to document representation and ranking based on interrelated entities.

This progression in research inspired a shift towards methodologies that treated entities not just as a latent layer but as explicit, integral elements of the retrieval model. For example, the creation of entity-based language models marked a significant development. \citet{raviv2016document} explored the impact of explicit entity markup within queries and documents, balancing term-based and entity-based information for document ranking. \citet{ensan2017document} developed the Semantic Enabled Language Model, which ranks documents based on semantic relatedness to the query.

Xiong et al.'s line of work \cite{xiong2017jointsem,xiong2017word,xiong2018towards,xiong2017explicit} exemplifies a dual-layered approach that pairs a traditional bag of terms representation with a separate bag of entities representation, enhancing the document retrieval process by incorporating both term and entity-based semantics. For example, Explicit Semantic Ranking used a knowledge graph to create "soft matches" in the entity space, and the Word-Entity Duet Model captured multiple interactions between queries and documents using a mixture of term and entity vectors.

The Entity-Duet Ranking Model (EDRM) \cite{liu-etal-2018-entity} represents a pioneering effort in neural entity-based search, merging the word-entity duet framework with the capabilities of neural networks and knowledge graphs (KGs). \citet{tran2022dense} advanced this area by introducing a method that clusters entities within documents to produce multiple entity ``views'' or perspectives,  enhancing the understanding and interpretation of various facets of a document. Recently, \citet{chatterjee2024dreq} proposed to learn query-specific weights for entities within candidate documents to re-rank them.

\paragraph{Entity ranking.} The task of entity ranking involves retrieving and ordering entities from a knowledge graph based on their relevance to a given query. Traditionally, this process has utilized term-based representations or descriptions derived from unstructured sources or structured knowledge bases like DBpedia \cite{lehmann2015dbpedia}. Ranking was commonly performed using models such as BM25 \cite{robertson2009probabilistic}. Additionally, Markov Randon Fields-based models like the Sequential Dependence Model \cite{metzler2005mrf} and its variants \cite{zhiltsov2015fielded,nikolaev2016paramterized,hasibi2016exploiting,raviv2012aranking} addressed the joint distribution of entity terms from semi-structured data.

As the availability of large-scale knowledge graphs increased, semantically enriched models were developed. These models leverage aspects such as entity types \cite{kaptein2010entity,balog2011query,garigliotti2017on} and the relationships between entities \cite{tonon2012combining,ciglan2012the} to enhance ranking accuracy.

More recently, the focus has shifted towards Learning-To-Rank (LTR) methods \cite{schuhmacher2015ranking,graus2016dynamic,dietz2019ent,chatterjee2021fine}, which utilize a variety of features, particularly textual information and neighboring relationships, to re-rank entities. The introduction of graph embedding-based models like GEEER \cite{gerritse2020graph} and KEWER \cite{nikolaev2020joint} has further enriched the field by incorporating Wikipedia2Vec \cite{yamada-etal-2020-wikipedia2vec} embeddings, allowing entities and words to be jointly embedded in the same vector space.

The latest advancements in this domain have been driven by transformer-based neural models such as GENRE \cite{cao2021autoregressive}, BERT-ER++ \cite{chatterjee2022berter}, and EM-BERT \cite{gerritse2022embert}. These models introduce sophisticated techniques including autoregressive entity ranking, blending BERT-based entity rankings with additional features, and augmenting BERT~\cite{Devlin2019BERTPO} with Wikipedia2Vec embeddings.

\section{Methodology}
In this section, we describe our approach to producing sparse representations of queries and documents that contain both entities and terms from a word piece vocabulary.
To do so, we incorporate entities into the model's vocabulary through the use of a Dynamic Vocabulary head.

\subsection{Sparse Encoders}
Given a query $q$ and a document $d$ as input, an LSR system uses a query encoder $f_q$ and a document encoder $f_d$ to convert the inputs into respective sparse representations $s_q$ and $s_d$. The dimensions are aligned with a vocabulary $\mathcal{V}$ and only a small number of dimensions have non-zero values.  Each dimension $s_q^i$ or $s_d^i$ encodes the weight of the $i^{th}$ vocabulary item ($v_i$) in the input query or document, respectively. The similarity between a query and a document is computed as the the dot product between the two corresponding sparse vectors:
\begin{equation}
    \small
    \mathcal{S}(q,d) = f_q(q)\cdot  f_d(d) = s_q \cdot s_d = \sum_{i=0}^{|\mathcal{V}|-1} s_q^i s_d^i
\end{equation}
Various types of sparse encoders have been previously defined in the literature and summarized by \citet{nguyen2023unified}. SPLADE~\cite{formal2021splade, formal2022distillation, lassance2022efficiency} is a state-of-the-art LSR method that employs the MLM architecture for both the query and document encoders. The strength of the MLM architecture is its ability to do term weighting and expansion in an end-to-end fashion, meaning that the model can itself learn from data to expand the input to semantically relevant terms and to weight the importance of individual terms. With an MLM encoder, the sparse representation of a query or document are generated as follows: 
\begin{equation}
    \small
    s_{(.)}^i  =\max_{ 0 \leq j < \mathcal{L}} log(1 + ReLU( e_i \cdot h_j))
    \label{eq:mlm}
\end{equation}
where $s_{(.)}{}^i$ and  $_i$ are the output weight and the embedding (from the embedding layer) of the $i^{th}$ vocabulary item respectively, $\mathcal{L}$ is the length of the input query or document, and $h_j$ is the the last hidden state of the $j^th$ query or document input token produced by a transformer backbone, such as BERT~\cite{Devlin2019BERTPO}.
A recent study \cite{nguyen2023unified} found that it is not necessary to have both query and document expansion. Disabling query expansion,by replacing the MLM query encoder by an MLP encoder can improve model efficiency while keeping the model's effectiveness. The MLP encoder weights each query input token as follows: 
\begin{equation}
    \small
    s_{q}^i = \sum_{0 \leq j < \mathcal{L}} \mathbbm{1}_{v_i = q_j} (\mathcal{W}\cdot h_j^T + b)
    \label{eq:mlp}
\end{equation}
where $\mathcal{W}$ and $b$ are parameters of a linear layer projecting a hidden state $h_j$ to a scalar weight. 

In this work, we employ this model variant with a MLP query encoder and a MLM document encoder as the baseline, and try to improve the model's expressiveness by expanding the output vocabulary to Wikipedia entities.
This model variant is similar to EPIC~\cite{macavaney2020expansion} and SPLADE-v3-Lexical~\cite{lassance2024splade}, though it does not exactly correspond to either model.
We call this model LSR-w to emphasize its use of the word piece vocabulary.

\begin{figure}[ht!]
    \centering
    \includegraphics[trim={7.5cm 4.5cm 8.9cm 4.5cm},clip,width=1.0\linewidth,]{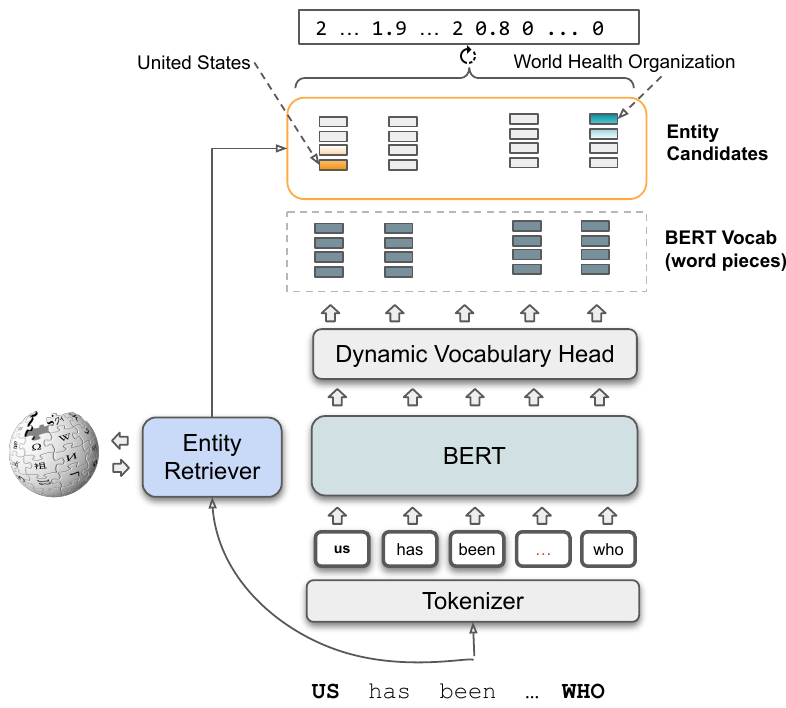}
    \caption{DyVo model with large entity vocabulary. The DyVo head scores entity candidates from an Entity Retriever component.}
    \label{fig:lsr-entity-arch}
    \vspace{-0.3cm}
\end{figure}

\subsection{Entity Vocabulary}\label{lsr_entity}
In this section, we describe our methodology to enrich the LSR vocabulary with Wikipedia entities. We build upon the MLM architecture for entity scoring in order to expand the input to any relevant items in the vocabulary, including entities which are not part of the encoder input. In the MLM head, the weight of the $i$-th entity with regard to an input query or document is calculated as follows: 
\begin{equation} \label{eq:mlm_entity}
    \small
    s_{ent}^i  =\lambda_{ent} \max_{ 0 \leq j < \mathcal{L}} log(1 + ReLU( e_i^{entity} \cdot h_j))
\end{equation}
We calculate the dot product between the entity embedding $e_i^{entity}$ and every hidden state $h_j$, and then select the maximum score. Via a ReLU gate, only positive weights are retained and then log scaled. For each query or document, only a small number of relevant entities have non-zero weights, forming a small bag of weighted entities (i.e., a sparse entity representation). This resulting entity representation is merged with the bag of words representation in the previous section to form a joint word-entity sparse representation. We add $\lambda_{ent}$ (initialized as 0.05) as a trainable scaling factor to adjust the entity weights. This scaling factor is important to prevent training collapse as discussed in Appendix~\ref{sec:collapse}

The final relevance score, which integrates both word and entity vocabularies, is computed as follows:
\begin{equation} \small S(q, d) = \sum_{i=0}^{|V|-1}s_w^i (q)s_w^i(d) + \sum_{j=0}^{|E|-1} s_{ent}^j(q) s_{ent}^j(d) \end{equation}

where $s_w^i(.)$ represents the weight of word $v^i$ and  $s_{ent}^j(.)$ represents the weight of entity $e^j$ with regard to the input query or document.

\subsection{Dynamic Vocabulary Head}
It is not practical to add every entity to the existing MLM head, because the MLM head exhaustively scores every term in its vocabulary for each input vector.
We propose a Dynamic Vocabulary (DyVo) head that augments an existing vocabulary using two ingredients: (1) embeddings of the new vocabulary terms (e.g., entity embeddings obtained from an external source) and (2) a candidate retrieval method that takes a query or document as input and identifies a small subset of the new vocabulary that may be present in the input (e.g., entities identified by an entity linker).
We use a DyVo head to expand the sparse encoder's vocabulary to include millions of Wikipedia entities, without the need to exhaustively score them as in Equation \ref{eq:mlm_entity}.

\subsubsection{Entity embeddings.} To produce a score for an entity in the vocabulary, the DyVo head needs to compute the dot product between the entity embedding and the hidden state of each input token. This operation requires both the entity embedding and the hidden states in the transformer backbone to have the same size and live in the same latent space. In this work, we chose DistilBERT~\cite{sanh2019distilbert}, which has proven its effectiveness in previous research, as the transformer backbone with an embedding size of 768. For our default entity embeddings, we utilize the \textbf{LaQue} pretrained dense entity encoder~\cite{arabzadeh2024laque} to encode entity descriptions from KILT~\cite{petroni-etal-2021-kilt} into entity embeddings. We choose LaQue for its consistent performance in yielding good entity weights and retrieval effectiveness in pilot experiments. We later provide detailed results comparing different types of entity embeddings. 

\subsubsection{Entity candidate retrieval.}  Instead of computing the weights for millions of entities in the vocabulary, we propose to add an entity candidate retrieval component (Figure \ref{fig:lsr-entity-arch}) that aims to narrow down the search space to a small set of relevant entities, which are then scored by the \ac{LSR} encoder using Equation \ref{eq:mlm_entity}. Offloading the entity retrieval task to a separate specialized component would allow the \ac{LSR} model to focus entirely on the scoring task to maximize the document retrieval objective. While using linked entities is a popular option in prior research, this approach may overlook important entities that are not directly mentioned in the text. Instead, we introduce a few-shot generative approach that leverages the power of LLMs to generate high quality candidates, including both linked entities and relevant entities. For each query, we show two examples and prompt LLMs (Mixtral, GPT4) to generate a list of Wikipedia entities that are helpful to retrieve relevant documents. The prompt template is shown in Prompt \ref{prompt_entity_retrieval}. We later compare our generative approach to various baselines. 

\subsubsection{Practical considerations.}
The DyVo head is memory-efficient when handling a large vocabulary, such as Wikipedia entities.
At both training time and inference time, DyVo avoids instantiating sparse vectors with millions of dimensions, which would require a substantial amount of memory compared to the raw text (e.g., 10MB to store a single float16 vector with 5 million dimensions).
During training, DyVo leverages the fact that the vast majority of entities do not appear in any given query (or document) to create a compact subset of the vocabulary for each batch. To do so, DyVo maintains a per-batch tensor of entity candidate IDs along with the corresponding entity weights, which are used to match entities between the query and the document. The weights of the matching entities are multiplied together and summed to produce the final relevance score.
This allows DyVo to instantiate relatively small sparse vectors that contain enough dimensions to hold the entity candidates, rather than instantiating vectors that correspond to the entire vocabulary.
Sparse representations are stored in an inverted index that is queried at inference time, so vocabulary-size vectors do not need to be instantiated at retrieval time. 
\section{Experimental setup}
\subsubsection{Datasets.} 
Given our need for entity-rich queries and documents, we evaluate our approach using datasets containing a mix of news documents and complex information needs (i.e., TREC Robust04, TREC Core 2018, CODEC), which have also been commonly used in prior work, e.g.~\cite{dalton2014entity, chatterjee2024dreq, tan2023incorporating, macavaney2019cedr, nogueira-etal-2020-document, li2023parade}. \textbf{Robust04}~\cite{voorhees2003overview} has 528k documents and 250 query topics where documents are news articles.
All topics are deeply annotated with 1246 judged documents per topic on average. \textbf{Core 2018} contains 595k news articles or blog posts from The Washington Post with about 50 topics and 524 relevant judgements per topic. \textbf{CODEC}~\cite{mackie2022codec} provides 729k web documents crawled from various sources and 42 complex query topics, covering recent themes (e.g., bitcoins, NFT) from diverse domains, such as history, economics, politics. Furthermore, each topic comes with approximately 147 document judgements and 269 entity annotations.

We use all provided topics (\textit{description} field on TREC datasets and \textit{query} field on CODEC) for evaluation. To train models, we used synthesized dataset provided by InParsV2~\cite{inparsv2}. Because CODEC is not available on InParsV2, we generate 10k queries ourselves using \textit{Mixtral-8x7B-Instruct-v0.1}~\cite{jiang2024mixtral}.

\subsubsection{Knowledge base and entity candidates.} We use the KILT~\cite{petroni-etal-2021-kilt} knowledge repository with 5.9 millions entities and only keep entities appearing in Wikipedia2Vec~\cite{yamada-etal-2020-wikipedia2vec}, resulting in \textasciitilde5.3 millions entities. To obtain linked entity candidates for queries, we use the REL~\cite{van2020rel} entity linker with n-gram NER tagger. For the entity retrieval approach, we experimented with different aproaches, including traditional sparse retrieval (BM25), dense retrieval (LaQue), and generative retrievers (Mixtral and GPT4). For BM25, we index the entity's description and retrieve the top 20 entities per query using Pyserini~\cite{pyserini_jimmy} with the default parameters. With LaQue, we encode both queries and entity descriptions using the LaQue (DistilBERT) dense encoder, and select the top 20 entities that have the highest dot product with the query's dense vector. With generative approaches, we prompt Mixtral and GPT4 to generate relevant entities and remove out-of-vocabulary entities. For simplicity, we re-use the linked entities from \citet{chatterjee2024dreq} on the document side for all experiments.

\subsubsection{Training configuration.} Starting from a LSR checkpoint without entities trained on MSMARCO, we further fine-tune them on the three datasets using the synthesized queries, MonoT5-3b scores for distillation, KL loss~\cite{formal2022distillation} and BM25 negatives. To regularize vector sparsity, we apply a L1 penalty on the output sparse representations, which has previously been shown to be effective~\cite{nguyen2023unified}.  We experiment with different L1 weights, including [1e-3, 1e-4, 1e-5]. For each setting, we train two LSR versions: LSR-w that produces word piece representations only, and DyVo that produces joint word-entity representations. On each dataset, we train the models for 100k steps with a batch size of 16, learning rate of 5e-7, and 16-bit precision on a single A100 GPU. Entity embeddings are pre-computed and frozen during training; only a projection layer is trained where the word and entity embedding sizes differ. 
\subsubsection{Evaluation metrics} We report commonly used IR metrics, including nDCG@10, nDCG@20 and R@1000 on all three datasets using the \textit{ir\_measures} toolkit~\cite{macavaney2022streamlining}. 

\section{Experimental results}
We first consider whether incorporating linked entities in sparse representations increases effectiveness over representations containing only word pieces, finding that doing so yields consistent improvements on our entity-rich benchmarks.
We then consider the impact of the entity selection component and the entity embeddings used, finding that performing entity retrieval rather than entity linking can further improve performance and that DyVo performs well with a range of entity embedding techniques.

\subsection*{RQ1: Does incorporating linked entities improve the effectiveness of LSR?} 
In this RQ, we seek to evaluate the effectiveness of \ac{LSR} with linked entities. We train three different \ac{LSR} versions with different sparse regularization weights (1e-3, 1e-4, 1e-5). For each \ac{LSR} version, we trained two models (LSR-w and DyVo) with and without entities, respectively, using exactly the same training configuration. Although we are mainly interested in the comparison between DyVo and LSR-w, other baselines (e.g., BM25, BM25+RM3, and zero-shot single-vector dense retrieval methods) are provided in Table \ref{tab:lsr-linked-entities} to help readers position LSR with regard to other first-stage retrieval families.

Our first observation is that our model with linked entities (DyVo) outperforms the model without entities (LSR-w) consistently on all metrics (nDCG@10, nDCG@20, R@1000) across three different datasets and three different sparsity constraints.  The difference between the two models is more pronounced when the document representations become more sparse. With the largest regularization weight (reg=1e-3), the documents are the most sparse and have the fewest terms. In this scenario, enriching the word representation with linked entities typically results in a significant gain, notably with an increase ranging from 1.15 to 3.57 points in nDCG@10 across all datasets. When we relax the sparsity regularization to 1e-4 and 1e-5, we observe an improvement in the performance of LSR-w baseline models. However, we still consistently observe the usefulness of linked entities, albeit to a lesser degree. In the most relaxed setup (reg=1e-5), we often gain from 1 to 2 nDCG points on all three datasets. The R@1000 improvement is similar, except we only observe a minimal increase on Core 2018.

Compared to other families, both LSR and DyVo demonstrate better performance than unsupervised lexical retrieval methods (BM25, BM25+RM3) and \ac{DR} models, including DistilBERT-dot-v5~\cite{reimers2019sentence}, GTR-T5-base~\cite{ni-etal-2022-large}, and Sentence-T5-base~\cite{ni-etal-2022-sentence}. Despite using models three times larger (T5-base vs. DistilBERT), both GTR-T5-base and Sentence-T5-base still show lower effectiveness than LSR models.
This is due to the generalization difficulties of dense retrieval methods.

DyVo also outperforms BM25 + RM3, a traditional query expansion method using pseudo-relevance feedback. Compared to GRF, a LLM-based query expansion approach by \citet{mackie2023generative}, DyVo achieves a significantly higher nDCG@10 score (e.g., 53.40 with DyVo using the REL linker versus 40.50 with GRF on CODEC). It is important to note that the LLMs used in GRF were not fine-tuned, and doing so would present substantial computational challenges.

\begin{table*}[ht!]
    \centering
    \small
    \begin{tabular}{lcccccccccc}
    \toprule
    \toprule
    \multirow{2}{*}{\textbf{Method}} & \multirow{2}{*}{\textbf{Reg}} &  \multicolumn{3}{c}{\textbf{TREC Robust04}} & \multicolumn{3}{c}{\textbf{TREC Core 2018}}  & \multicolumn{3}{c}{\textbf{CODEC}} \\
    \cline{3-5} \cline{6-8} \cline{9-11} 
     &  & \textbf{\tiny nDCG@10} & \textbf{\tiny nDCG@20} & \textbf{\tiny R@1k} & \textbf{\tiny nDCG@10} & \textbf{\tiny nDCG@20} & \textbf{\tiny R@1k} & \textbf{\tiny nDCG@10} & \textbf{\tiny nDCG@20} & \textbf{\tiny R@1k} \\ 
    \midrule 
    \multicolumn{10}{l}{\textit{Unsupervised sparse retrieval}} \\
    \midrule
    BM25  &  & 39.71 & 36.25 & 57.18 & 30.94 & 29.19 & 52.19 & 37.70 & 35.28 & 61.25 \\ 
    BM25 + RM3  & & 43.77 & 40.64 & 64.21 & 35.82 & 34.79 & 60.09 & 39.93 & 39.96 & 65.70 \\ 
    \midrule 
    \multicolumn{10}{l}{\textit{Zero-shot Dense Retrieval}}\\
    \midrule 
    DistilBERT-dot-v5 & & 37.95 & 34.97 & 52.41 & 37.02 & 34.60 & 54.07 & 42.76 & 46.67 & 60.33 \\
    GTR-T5-base  & & 43.79 & 39.33 & 54.35 & 38.81 & 36.51 & 57.62 & 48.42 & 54.01 & 66.96\\
    Sentence-T5-base  &  & 44.06 & 39.60 & 57.64 & 43.18 & 39.54 & 60.88 & 44.22 & 32.10 & 65.48\\ 
    \midrule
    \multicolumn{11}{l}{\textit{Learned Sparse Retrieval}}\\
    \midrule 
    LSR-w& \multirow{2}{*}{1e-3} & 40.37 & 37.23 & 55.66 & 34.50 & 31.45 & 52.66  & 39.10 & 35.32  & 57.58   \\
    DyVo (REL) & & \textbf{41.52} & \textbf{38.62} & \textbf{56.78} & \textbf{37.50} & \textbf{34.61} & \textbf{54.14} &  \textbf{42.67} & \textbf{38.32} & \textbf{59.81} \\
    \midrule
    LSR-w & \multirow{2}{*}{1e-4} & 47.69 & 44.48 & 64.47 & 38.94 & 37.37 & 60.44 & 50.54 & 46.71 & 66.39  \\
    DyVo (REL) &  & \textbf{48.15} & \textbf{44.85} & \textbf{64.72} & \textbf{43.10}  & \textbf{39.46} & \textbf{60.43} & \textbf{51.66} & \textbf{47.95} & \textbf{68.49} \\
    \midrule
    LSR-w & \multirow{2}{*}{1e-5} & 49.13 & 46.34 & 66.86 & 40.99 & 38.73 & 63.22 & 52.61 & 49.22 & 69.07  \\
    DyVo (REL) & & \textbf{51.19} & \textbf{47.65} & \textbf{68.56} & \textbf{43.72} & \textbf{40.56} & \textbf{63.56} & \textbf{53.40} & \textbf{51.15} & \textbf{70.60} \\
    \bottomrule
    \bottomrule
    \end{tabular}
    \caption{Results with linked entities. All LSR models use a DistilBERT backbone. DyVo uses entities found by the REL entity linker and LaQue entity embeddings. All documents are truncated to the first 512 tokens.}
    \label{tab:lsr-linked-entities}
    \vspace{-0.3cm}
\end{table*}

\subsection*{RQ2: Can LSR be more effective with retrieval-oriented entity candidates?}
In the previous RQ, we explored how incorporating linked entities enhances \ac{LSR}'s representations. However, relying solely on linked entities overlooks other relevant entities crucial for document retrieval. For instance, with the CODEC query ``Why are many commentators arguing NFTs are the next big investment category?'', entities like ``Cryptocurrency'', ``Bitcoin'', and ``Digital asset'' can be valuable despite not being explicitly mentioned.

In this RQ, we aim to evaluate our few-shot generative entity retrieval approach based on Mixtral or GPT4 and compare it with other entity retrieval approaches, including entity linking (as explored in the previous RQ), sparse methods (BM25), dense entity retrieval methods (LaQue), and human annotations. The results are shown in Table \ref{tab:relevant-entities}.

Observing the table, we note that DyVo (BM25) and DyVo (LaQue) show modest performance gains compared to the DyVo (REL) model, which incorporates linked entities, and the LSR model without entities. Employing LaQue-retrieved candidates to DyVo increases LSR-w's  R@1000 by +1.39 points (66.86 $\rightarrow$ 68.25), +1.61 points (63.22 $\rightarrow$ 64.83), and +1.8 points (from 69.07 $\rightarrow$ 70.87) on the Robust04, Core18, and CODEC datasets, respectively. This recall improvement is comparable to the gain achieved by using the REL entity linker. However, we generally observe no benefits in terms of nDCG when using the BM25 or LaQue retriever. This could be because BM25 and LaQue tend to prioritize recall, resulting in the retrieval of not only relevant entities but also noisy entities. 
\begin{table*}[ht!]
    \centering
    \small
    \begin{tabular}{lccccccccc}
    \toprule
    \toprule
    \multirow{2}{*}{\textbf{Method}}  &  \multicolumn{3}{c}{\textbf{TREC Robust04}} & \multicolumn{3}{c}{\textbf{TREC Core 2018}}  & \multicolumn{3}{c}{\textbf{CODEC}} \\
    \cline{2-4} \cline{5-7} \cline{8-10} 
     &  \textbf{\tiny nDCG@10} & \textbf{\tiny nDCG@20} & \textbf{\tiny R@1k} & \textbf{\tiny nDCG@10} & \textbf{\tiny nDCG@20} & \textbf{\tiny R@1k} & \textbf{\tiny nDCG@10} & \textbf{\tiny nDCG@20} & \textbf{\tiny R@1k} \\ 
    \midrule
    LSR-w & 49.13 & 46.34 & 66.86 & 40.99 & 38.73 & 63.22 & 52.61 & 49.22 & 69.07  \\
    DyVo (REL)  & 51.19 & 47.65 & 68.56 & 43.72 & 40.56 & 63.56 & 53.40 & 51.15 & 70.60 \\
    DyVo (BM25)  & 51.38  & 47.72 & 67.74 & 42.48 & 38.89 & 64.58 & 53.25 & 49.80 & 69.83 \\
    DyVo (LaQue) & 49.42 & 46.31 & 68.25 & 40.24 & 38.39 & 64.83 & 53.73 & 50.34 & 70.87 \\
    DyVo (Mixtral)  & 52.97 & 49.21 & 69.28 & \textbf{43.80} & 41.86 & 68.27 & 54.90 & 52.82 & 73.20  \\
    DyVo (GPT4)  & \textbf{54.39} & \textbf{50.89} & \textbf{70.86} & 43.06 & \textbf{42.25} & \textbf{68.57} & \textbf{56.46} & \textbf{53.72} & 74.47 \\
    DyVo (Human)  & - & - & - & - & - & - &  56.42 & 52.96 & \textbf{75.13}  \\
    \bottomrule
    \bottomrule
    \end{tabular}
    \caption{Results with entities retrieved by different retrievers. All models are trained with a DistilBERT backbone, LaQue entity embeddings, and L1 regularization (weight=1e-5).}
    \vspace{-0.3cm}
    \label{tab:relevant-entities}
\end{table*}

\begin{table*}[ht]
    \centering
    \small
    \begin{tabular}{llccccccccccc}
    \toprule
    \toprule
    \multirow{2}{*}{\textbf{Method}}  & \multirow{2}{*}{\textbf{Entity Rep.}} &  \multicolumn{3}{c}{\textbf{TREC Robust04}} & \multicolumn{3}{c}{\textbf{TREC Core 2018}}  & \multicolumn{3}{c}{\textbf{CODEC}}\\
    \cline{3-5} \cline{6-8} \cline{9-11} 
     & & \textbf{\tiny nDCG@10} & \textbf{\tiny nDCG@20} & \textbf{\tiny R@1k} & \textbf{\tiny nDCG@10} & \textbf{\tiny nDCG@20} & \textbf{\tiny R@1k} & \textbf{\tiny nDCG@10} & \textbf{\tiny nDCG@20} & \textbf{\tiny R@1k}\\ 
    \midrule
    \midrule
    LSR-w &  -  & 49.13 & 46.34 & 66.86 & 40.99 & 38.73 & 63.22 & 52.61 & 49.22 & 69.07  \\
    DyVo (GPT4) & Token Aggr.  & 51.35& 48.01& 67.46& 41.63& 39.37& 64.01& 53.44& 50.39& 69.94\\
    DyVo (GPT4) & DPR & 48.68& 45.77& \textbf{75.21} & 40.26& 37.52& \textbf{70.81} & 53.04& 49.18& \textbf{75.19} \\
    DyVo (GPT4) & JDS  & 51.21& 48.38& 73.79& 44.29& 41.86& 70.16& 55.08& 50.93& 73.97 \\ 
    DyVo (GPT4) & Wiki2Vec & 54.04& 50.21& 69.85& 44.15& \textbf{43.13} & 67.77& 56.30& 53.25& 73.03\\
    DyVo (GPT4) & LaQue & 54.39& 50.89& 70.86& 43.06& 42.25& 68.57& 56.46& 53.72& 74.47\\
    DyVo (GPT4) & BLINK & \textbf{55.56} & \textbf{51.71} & 71.81& \textbf{44.63} & 42.94& 69.11& \textbf{58.15} & \textbf{54.83} & 74.72\\
    \bottomrule
    \bottomrule
    \end{tabular}
    \caption{Results with different entity embeddings.  All models are trained with a DistilBERT backbone and L1 regularization (weight=1e-5).  Entity candidates generated by GPT4 are used on queries for inference.}
    \label{tab:ablation_entity_rep}
    \vspace{-0.3cm}
\end{table*}

Our generative approach utilizing Mixtral and GPT4 represents a significant step forward in entity retrieval for document ranking. Compared to linked entities provided by REL, our approach showcases notable improvements, enhancing nDCG@10 and nDCG@20 scores by approximately +1.3 to +1.78 points across all datasets, with the exception of nDCG@10 on Core 2018. Mixtral's effectiveness is further highlighted by its impact on R@1000 scores, with increases observed across the Robust04, Core 2018, and CODEC datasets.

Additionally, when we replace Mixtral with GPT4, we see further improvements that result in DyVo achieving the highest performance on nearly every metric and dataset.
Notably, retrieval using GPT-4 generated entities is competitive with retrieval using human-annotated entities on CODEC, underlining the significance of enriching query representations with relevant entities beyond linked ones. We attach examples in Table~\ref{tab:entity_candidate_example} in the Appendix to illustrate the candidates retrieved by different systems.

\subsection*{RQ3: How does changing entity embeddings affect the model's ranking performance?}
Previously, we utilized the same entity encoder, LaQue~\cite{arabzadeh2024laque}, to generate entity embeddings. Here, our objective is to evaluate various approaches to obtain entity embeddings including Token Aggregation (i.e., splitting an entity's surface form into word pieces and averaging their static embeddings), Wikipedia2Vec~\cite{yamada-etal-2020-wikipedia2vec}, general dense passage encoders like JDS and DPR~\cite{pouran-ben-veyseh-etal-2021-dpr} and specialized dense entity encoders like LaQue and BLINK~\cite{arabzadeh2024laque,laskar-etal-2022-blink}. JDS is a joint dense ([CLS] vector) and sparse model with a shared DistilBERT backbone. We train our JDS model on MSMARCO dataset with a dual dense-sparse loss, using it to encode entity descriptions into dense embeddings. The results is shown in Table \ref{tab:ablation_entity_rep}.

First, we observe that simply tokenizing the entity name into word pieces and averaging the transformer's static token embeddings proves to be a viable method for creating entity embeddings. This approach typically yields a +1 point improvement over LSR-w across various metrics and datasets. We hypothesize that this improvement mainly stems from phrase matching through entity name matching, as we believe the token static embeddings do not encode much entity knowledge.

Interestingly, in terms of nDCG scores, this simple method outperforms the DPR and JDS methods, which rely on generic dense passage encoders trained for ad-hoc passage retrieval tasks to encode entity descriptions. DPR and JDS, however, demonstrate strong recall, suggesting that these encoders may prioritize encoding abstract entity information, which enables them to pull relevant documents within the top 1000 results. However, they may lack fine-grained entity knowledge necessary for more nuanced weighting.

Wikipedia2Vec (Wiki2Vec, dim=300), LaQue, and BLINK are specialized for entity representation learning or entity ranking tasks. As indicated in the last three rows of Table \ref{tab:ablation_entity_rep}, using them to generate entity embeddings enhances document retrieval performance across all metrics and datasets. Despite being trained using a simple skip-gram model, Wikipedia2Vec effectively supports \ac{LSR} in document retrieval, outperforming models utilizing aggregated token embeddings and dense passage encoders. The robustness of Wikipedia2Vec has been documented in prior research~\cite{oza2023entity}. Substituting Wikipedia2Vec with more advanced transformer-based entity encoders such as LaQue and BLINK results in the strongest overall performance. LaQue, based on the lightweight DistilBERT backbone, shows a slight improvement over Wikipedia2Vec. Using a larger transformer model (BERT-large), BLINK usually achieves a +1 nDCG point increase compared to LaQue, topping all datasets in terms of nDCG@10 and nDCG@20.

\section{Conclusion}
LSR has emerged as a competitive method for first-stage retrieval. In this work, we observed that relying on only word pieces for lexical grounding can create ambiguity in sparse representations--- especially when entities are split into subwords. We explored whether learned sparse representations can include entity dimensions in addition to word piece dimensions. In order to facilitate modeling millions of potential entities, we proposed a Dynamic Vocabulary (DyVo) head that leverages entity retrieval to identify potential entity candidates and entity embeddings to represent them.
We find that while both linked entities and LLM-generated entities are effective, LLM-generated entities ultimately yield higher LSR effectiveness. The approach is largely robust to the choice of entity embedding. Our work sets the stage for other LSR models that go beyond word piece vocabularies.

\section*{Limitations}
While our approach is highly effective on the document retrieval benchmarks considered, it is important to note that its reliance on large language models (LLMs) like Mixtral and GPT4 can pose computational and cost inefficiencies. This challenge is not unique to our methodology; rather, it is a common concern across various research pursuits employing LLMs for retrieval purposes. One potential avenue for mitigating these costs involves leveraging LLMs to generate synthetic datasets and distill their internal knowledge into a more streamlined entity ranker or re-ranker. Addressing this issue extends beyond the scope of our current work.

\section*{Ethics Statement}
We constructed our LSR encoder using a pretrained DistilBERT and employed Large Language Models such as Mixtral and GPT4 to generate entity candidates. Consequently, our models may inherit biases (e.g., preferences towards certain entities) encoded within these language models. Our evaluation encompasses both open-source models (Mixtral, DistilBERT, LaQue, BLINK, REL, Wikipedia2Vec) and proprietary ones (GPT4), which do not always disclose their training data.

\section*{Acknowledgements}
This research was supported by the Hybrid Intelligence Center, a 10-year program funded by the Dutch Ministry of Education, Culture and Science through the Netherlands Organisation for Scientific Research, \url{https://hybrid-intelligence-centre.nl}, and project VI.Vidi.223.166 of the NWO Talent Programme which is (partly) financed by the Dutch Research Council (NWO).

\bibliography{anthology,custom}
\bibliographystyle{acl_natbib}

\appendix

\section{Appendix}
\label{sec:appendix}
\begin{pabox}[float*=b, label={prompt_query_generation}, width=\textwidth]{Prompt template for query generation with LLMs}
  \textit{Given an input document, your task is to generate a short and self-contained question that could be answered by the document. Three examples are given, please finish generating the query for the last example. Please generate only one short and self-contained question without numbering in a single line, and do not generate an explanation.}\\
    \textbf{Example 1:}\\
    \textbf{Document:} We don’t know a lot about the effects of caffeine during pregnancy on you and your baby. So it‘s best to limit the amount you get each day. If you are pregnant, limit caffeine to 200 milligrams each day. This is about the amount in 1½ 8-ounce cups of coffee or one 12-ounce cup of coffee.\\
    \textbf{Relevant Query:} Is a little caffeine ok during pregnancy?\\
    \textbf{Example 2:}\\
    \textbf{Document:} Passiflora herbertiana. A rare passion fruit native to Australia. Fruits are green-skinned, white fleshed, with an unknown edible rating. Some sources list the fruit as edible, sweet and tasty, while others list the fruits as being bitter and inedible.assiflora herbertiana. A rare passion fruit native to Australia. Fruits are green-skinned, white fleshed, with an unknown edible rating. Some sources list the fruit as edible, sweet and tasty, while others list the fruits as being bitter and inedible.\\
    \textbf{Relevant Query:} What fruit is native to Australia?\\
    \textbf{Example 3:}\\
    \textbf{Document:} The Canadian Armed Forces. 1  The first large-scale Canadian peacekeeping mission started in Egypt on November 24, 1956. 2  There are approximately 65,000 Regular Force and 25,000 reservist members in the Canadian military. 3  In Canada, August 9 is designated as National Peacekeepers’ Day.\\
    \textbf{Relevant Query:} How large is the canadian military?\\
    \textbf{Example 4:}\\
    \textbf{Document:} \{input document\}\\
    \textbf{Relevant Query:};
  \end{pabox}

  \begin{pabox}[label={prompt_entity_retrieval},float*=b, width=\textwidth]{Prompt template for few-shot generative entity retrieval}
\textit{Identify Wikipedia entities that are helpful to retrieve documents relevant to a web search query. Please return a list of entity names only}: \\ 
    \textbf{Example 1:} \\ 
    \textbf{Query:} How is the push towards electric cars impacting the demand for raw materials?\\
    \textbf{Entities:} ["Cobalt", "Automotive battery", "China", "Electric car", "Electric battery", "Gigafactory 1", "Demand", "Fossil fuel", "Electric vehicle industry in China", "Electric vehicle battery", "Electric vehicle conversion", "Electric vehicle", "Supply and demand", "Mining industry of the Democratic Republic of the Congo", "Raw material", "Lithium iron phosphate", "Lithium-ion battery", "Mining", "Lithium", "Petroleum"] \\
    \textbf{Example 2:} \\
    \textbf{Query:}  Why do many economists argue against fixed exchange rates?\\
    \textbf{Entities:} ["Argentine peso", "Currency crisis", "Inflation", "Hong Kong dollar", "Exchange rate", "Gold standard", "European Exchange Rate Mechanism", "1998 Russian financial crisis", "Black Saturday (1983)", "Black Wednesday", "Optimum currency area", "Mexican peso crisis", "Milton Friedman", "Euro", "Recession", "Currency intervention", "1997 Asian financial crisis", "Devaluation", "Original sin (economics)", "Exchange-rate regime"]\\
    \textit{Please find relevant entities for this new example:}\\
    \textbf{Query:} \{input query\}\\
    \textbf{Entities:} 
\end{pabox}
\subsection{Detailed training configuration}
We train our DyVo methods using two-step distillation. In the first step, we train a base LSR model on MSMARCO without entities using standard LSR training techniques. We employ KL loss to distill knowledge from a cross-encoder with data obtained from \textit{sentence-transformers}~\cite{reimers-2019-sentence-bert}\footnote{https://huggingface.co/datasets/sentence-transformers/msmarco-hard-negatives}. This model is trained with a batch size of 64 triplets (query, positive passage, negative passage) for 300k steps with 16-bit precision. In the second step, we start from the model pretrained on MSMARCO and further fine-tune it on the target datasets using distillation training on synthesized queries, BM25 negatives, and cross-encoder scores from MonoT5-3B~\cite{nogueira-etal-2020-document}. The documents in Robust04, Core 2018, and CODEC are longer than MSMARCO, so we use a smaller batch size of 16. All models are trained on one single A100 for for 100k steps.

For query generation, we re-use generated queries from InParsv2~\cite{inparsv2} available for TREC Robust04 and Core 2018. For CODEC, we generate the queries ourselves using the prompting Mixtral model. We re-use the prompt template in InParsv2 and add a small instruction at the beginning (Prompt \ref{prompt_query_generation}).

For sparse regularization, we apply L1 with varying regularization strengths. Entity representations are sparse themselves since we constrain the output to a small set of entity candidates and ignore other entities. Therefore, we do not apply L1 to entities.

\begin{figure}[ht!]
    \centering
    \includegraphics[width=0.8\linewidth]{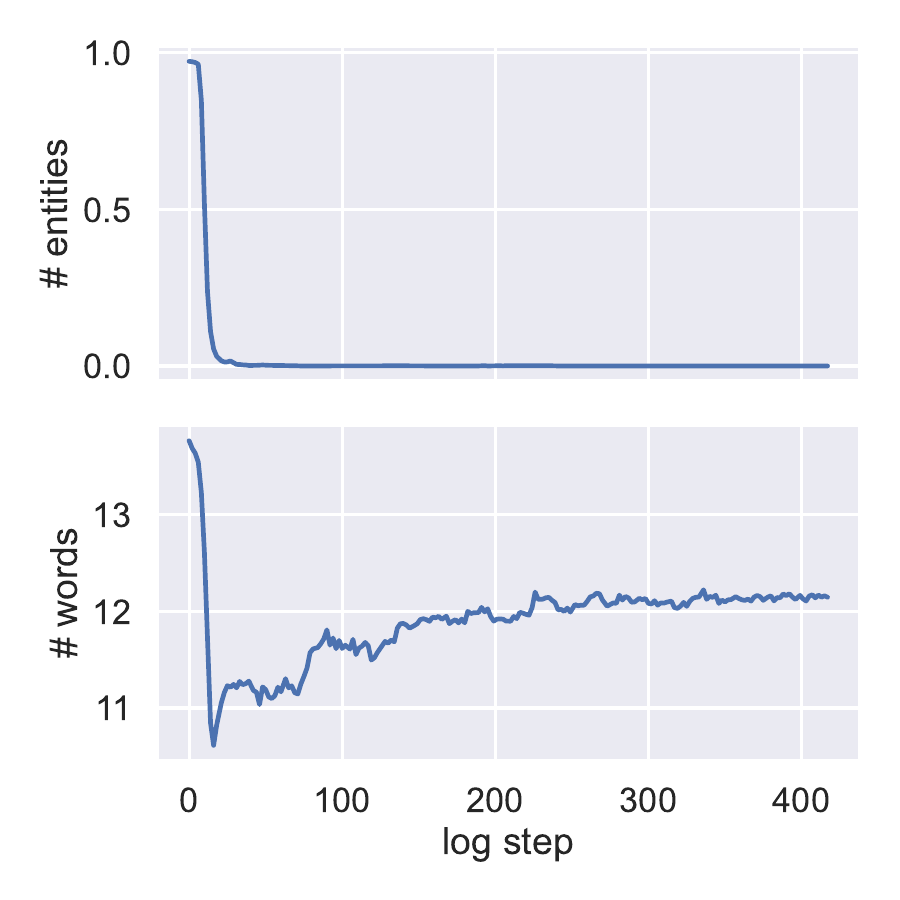} 
    \caption{Entity representation collapse during training.}
    \label{fig:training_collapse}
\end{figure}

\subsection{Entity representation collapse}
\label{sec:collapse}
When integrating entity embeddings into DyVo, we observe that the model produces entity weights with magnitudes significantly higher than those of word piece weights. This discrepancy may arise from the lack of alignment between entity embeddings, generated by a separate model, and word piece embeddings. Initially, the model attempts to mitigate the dominance of entity weights by scaling them down. However, after a certain number of training epochs, the model overcompensates, resulting in the collapse of entity representations. This collapse is illustrated in Figure \ref{fig:training_collapse}, where all entity weights become negative and are subsequently filtered out by the ReLU gate. Once this collapse occurs, it cannot be rectified, as there is no gradient flowing through the ReLU gate. To address this issue, we introduce a learnable scaling factor, as depicted in Equation \ref{eq:mlm_entity}, initializing it to a small value. This scaling factor is helpful to alleviate entity dominance at the beginning of training and temper the model's aggressiveness in scaling down entity weights during training.

\begin{table*}
    \centering
    \begin{tabular}{lp{13.4cm}}
    \toprule
    \toprule
        \textbf{Retriever} &  \textbf{Q}: \textit{``How vital was French support during the American Revolutionary War?''}\\
     & \textbf{WP} :  \small{[how, vital, was, french, support, during, the, american, revolutionary, war, ?]}\\
    \midrule
      REL   &  \small{[Vitalism, French people, Military logistics, American Revolutionary War]}
 \\
     BM25 & \small{[Richard Howe, 1st Earl Howe, HMS Childers (1778), Robert Howe (Continental Army officer), James Coutts Crawford, Glorious First of June, George Eyre, Jacques-Antoine de Chambarlhac de Laubespin, Anthony James Pye Molloy, Nantucket during the American Revolutionary War era, Friedrich Joseph, Count of Nauendorf, Jonathan Faulknor the elder, Joseph Spear, HMS Romney (1762), HMS Roebuck (1774), France in the American Revolutionary War, Invasion of Corsica (1794), List of British fencible regiments, Northern theater of the American Revolutionary War after Saratoga, Robert Linzee, Guilin Laurent Bizanet]
}\\ 
      LaQue & \small{[France in the American Revolutionary War, List of French units in the American Revolutionary War, Support our troops, List of wars involving France, List of American Revolutionary War battles, American Volunteers, Colonial American military history, List of battles involving France in modern history, Military history of France, List of the lengths of United States participation in wars, 1776, France and the American Civil War, USS Confederacy (1778), Financial costs of the American Revolutionary War, List of wars involving the United States, List of American Civil War generals (Union), United States assistance to Vietnam, French Revolutionary Wars, American Revolutionary War, List of battles involving France]} \\ 
      Mixtral & \small{[American Revolutionary War, France, United States, Military history, Diplomacy, Military alliance]
} \\ 
       GPT4 & \small{[France in the American Revolutionary War, French Army, American Revolutionary War, Benjamin Franklin, Kingdom of France, Treaty of Alliance (1778), George Washington, John Adams, Treaty of Paris (1783), Continental Congress, Continental Army, Naval battles of the American Revolutionary War, Siege of Savannah, Capture of Fort Ticond]
}\\
       Human & \small{[American Revolution, France in the American Revolutionary War, Kingdom of Great Britain, United States, George Washington, Roderigue Hortalez and Company, British Empire, France, George Washington in the American Revolution, Gilbert du Motier, Marquis de Lafayette, Spain and the American Revolutionary War, American Revolutionary War, Diplomacy in the American Revolutionary War, Treaty of Paris (1783), Franco-American alliance, Naval battles of the American Revolutionary War, Treaty of Alliance (1778), Battles of Saratoga]}
 \\
    \midrule
     &  \textbf{Q}: \textit{Why are many commentators arguing NFTs are the next big investment category?}\\
     & \textbf{WP}: \small{[why, are, many, commentators, arguing, \textcolor{red}{n}, \textcolor{red}{\#\#ft}, \textcolor{red}{\#\#s}, are, the, next, big, investment, category]} \\ 
    \midrule
      REL   &  \small{[Sports commentator, National Film and Television School, Next plc, Toronto, Investment banking, Categorization]}
 \\
     BM25 & \small{[Kuznets swing, The Green Bubble, Why We Get Fat, Big mama, Types of nationalism, Not for Tourists, Mark Roeder, Ernie Awards, Dramatistic pentad, Pagan Theology, RJ Balaji, Leslie Hardcastle, Why didn't you invest in Eastern Poland?, Big Data Maturity Model, Celebrity Big Brother racism controversy, The Bottom Billion, National Film and Television School, Canopy Group, The Wallypug of Why]}\\ 
      LaQue & \small{[List of bond market indices, National Futures Association, NB Global, Companies listed on the New York Stock Exchange (N), Companies listed on the New York Stock Exchange (G), Companies listed on the New York Stock Exchange (F), List of exchange-traded funds, Companies listed on the New York Stock Exchange (T), Emerging and growth-leading economies, List of private equity firms, List of wealthiest organizations, Pension investment in private equity, Group of Ten (economics), Companies listed on the New York Stock Exchange (P), List of stock market indices, Lists of corporate assets, Companies listed on the New York Stock Exchange (U), List of public corporations by market capitalization, Net capital outflow, National best bid and offer]} \\ 
      Mixtral & \small{[Non-fungible token, Blockchain, Cryptocurrency, Digital art, Ethereum, Value proposition, Art market, CryptoKitties, Investment strategy]} \\ 
       GPT4 & \small{[Non-fungible token, Cryptocurrency, Bitcoin, Ethereum, Digital art, Blockchain, CryptoKitties, Digital asset, Cryptocurrency bubble, Cryptocurrency exchange, Initial coin offering, Cryptocurrency wallet, Smart contract, Decentralized application, Digital currency]}\\
       Human & \small{[Cryptocurrency, Public key certificate, Blockchain, Virtual economy, Bitcoin, Speculation, Non-fungible token, Ethereum]}
 \\
    \bottomrule
    \bottomrule
    \end{tabular}
    \caption{Example of relevant entities retrieved by different systems. List of word pieces (\textbf{WP}) returned by DistilBERT tokenizer is  shown under each query.}
    \label{tab:entity_candidate_example}
\end{table*}

\subsection{Qualitative comparison of different entity retrieval systems}
In Table \ref{tab:entity_candidate_example}, we provide a qualitative comparison of the entity candidates retrieved by different systems. Within the two query samples presented, we observe that the generative approaches (i.e., Mixtral and GPT4) consistently produce highly relevant entities. Notably, Mixtral tends to generate fewer and shorter entities compared to both GPT-4 and human annotations. Conversely, GPT4 retrieves more entities, and sometimes more entities than human-produced candidates. This discrepancy suggests an explanation for why Mixtral's performance in generating entities to support document retrieval falls short of that achieved by GPT4.

In contrast to the consistent performance of generative entity retrieval, we observe divergent behaviors among other approaches (i.e., REL, BM25, and LaQue) across the two queries. The first query, which is less ambiguous with clearly expressed entities, allows these systems to retrieve/link simple, direct entities such as ``American Revolutionary War'' and ``France in the American Revolutionary War''. However, they also introduce a significant amount of noise with irrelevant entities.

Conversely, the second query poses greater difficulty, with the entity ``Non-fungible token'' mentioned via its abbreviation ``NFTs'' which is further fragmented by the DistilBERT tokenizer into meaningless sub-word units. In this scenario, REL and BM25 fail entirely, while LaQue manages to retrieve only generic and distantly relevant entities. None of these systems successfully resolves ``NFTs'' to ``Non-fungible token'' as the generative approach does.

\end{document}